\documentclass[utf8]{frontiersSCNS} 

\usepackage{url,lineno,microtype,subcaption}
\usepackage[onehalfspacing]{setspace}
\usepackage{graphicx}
\usepackage{mathrsfs}
\usepackage{amsmath}
\usepackage{natbib}
\usepackage{color}
\usepackage{changes}

\newcommand\bb[1]{\mbox{\boldmath{$#1$}}}

\def\keyFont{\fontsize{8}{11}\helveticabold }
\def\firstAuthorLast{Cerri et al.} 
\def\Authors{Silvio Sergio Cerri$^{1,*}$, Daniel Gro\v{s}elj$^{2}$, Luca Franci$^{3}$}
\def\Address{$^{1}$Department of Astrophysical Sciences, Princeton University, Princeton, NJ, USA;\\
$^{2}$Max-Planck-Institute for Plasma Physics, Garching, Germany\\
$^{3}$School of Physics and Astronomy, Queen Mary University of London, London, UK;}
\def\corrAuthor{Peyton Hall, 4 Ivy Ln, Princeton, NJ 08544, USA}
\def\corrEmail{scerri@astro.princeton.edu}

\begin{document}
\onecolumn
\firstpage{1}

\title[kinetic plasma turbulence]{Kinetic plasma turbulence: recent insights and open questions from 3D3V simulations} 

\author[\firstAuthorLast ]{\Authors} 
\address{} 
\correspondance{} 

\extraAuth{}

\maketitle

\begin{abstract}

Turbulence and kinetic processes in magnetized space plasmas have been extensively investigated over the past decades via \emph{in-situ} spacecraft measurements, theoretical models and numerical simulations. 
In particular, multi-point high-resolution measurements from the \emph{Cluster} and \emph{MMS} space missions brought to light an entire new world of processes, taking place at the plasma kinetic scales, and exposed new challenges for their theoretical interpretation.
A long-lasting debate concerns the nature of ion and electron scale fluctuations in solar-wind turbulence and their dissipation via collisionless plasma mechanisms.
Alongside observations, numerical simulations have always played a central role in providing a test ground for existing theories and models.  
In this Perspective, we discuss the advances achieved with our 3D3V (reduced and fully) kinetic simulations, as well as the main questions left open (or raised) by these studies. 
To this end, we combine data from our recent kinetic simulations of both freely decaying and continuously driven fluctuations to assess the similarities and/or differences in the properties of plasma turbulence in the sub-ion range. 
Finally, we discuss possible future directions in the field and highlight the need to combine different types of numerical and observational approaches to improve the understanding of turbulent space plasmas.

\tiny
 \keyFont{ \section{Keywords:} magnetic fields, plasma turbulence, solar wind, kinetic plasma simulations, turbulence intermittency, plasma waves}
\end{abstract}

\section{Introduction}\label{sec:intro}

With the establishment of satellite space missions, the near-Earth environment and the solar wind have provided unique opportunities to explore the physics of weakly collisional, magnetized plasmas~\citep[e.g.,][]{BrunoCarboneLRSP2013,ChenJPP2016,Verscharen2019}. 
In particular, increasingly accurate \emph{in-situ} measurements of plasma fluctuations and particle distribution functions from \emph{Cluster} and \emph{MMS} have uncovered an entire new world of kinetic processes occurring in plasma turbulence~\citep[e.g.,][]{AlexandrovaPRL2009,AlexandrovaAPJ2012,AlexandrovaSSRv2013,SahraouiPRL2009,SahraouiPRL2010,ChenPRL2010,ChenNATCo2019,GrecoAPJL2016,NaritaAPJL2016,ChasapisAPJL2017,ChenBoldyrevAPJ2017,HuangAPJL2017,RobertsAPJ2017,ServidioPRL2017}.
These observations highlight a change in the turbulent cascade at plasma microscales, challenging the community for a consistent theory of kinetic-range turbulence. In fact, several collisionless plasma processes may be simultaneously at play and compete with each other in determining the nature of ion-scale and electron-scale fluctuations~\citep[e.g.,][]{StawickiJGR2001,GaltierPOP2003,HowesJGR2008,GaryJGR2009,SchekochihinAPJS2009,HeAPJL2012,PodestaJGR2012,BoldyrevPerezAPJL2012,BoldyrevAPJ2013,MatthaeusAPJ2014,PassotSulemAPJL2015,PassotSulem_arxiv2019,KunzJPP2018,PassotPOP2018} and, consequently, how free energy cascades in phase space~\citep[e.g.,][]{SchekochihinPPCF2008,ServidioPRL2017,AdkinsSchekochihinJPP2018,CerriAPJL2018,EyinkPRX2018,PezziPOP2018,KawazuraPNAS2019}.
Many observations at ion and sub-ion scales, specifically, suggest that turbulent fluctuations exhibit properties mainly typical of kinetic Alfv\'en waves (KAWs)~\citep{LeamonJGR1998,SahraouiPRL2009,PodestaTenBargeJGR2012,SalemAPJL2012,ChenPRL2013,ChenJPP2016,KiyaniApJ2013,LacombeAPJ2017}.
The emergence of KAW-like fluctuations in kinetic turbulence has been also supported by means of a large number of theoretical and numerical works~\citep[e.g.,][]{HollwegJGR1999,StawickiJGR2001,GaryNishimuraJGR2004,HowesJGR2008,GaryJGR2009,SahraouiAPJ2012,TenBargeApJ2012,VasconezPOP2014,VasconezAPJ2015,FranciAPJ2015,CerriAPJL2016,PucciJGR2016,ZhaoJGR2016,ValentiniA&A2017,GroseljPRX2019}.
Some of these studies rely on the so-called spectral field ratios, which provide a measure of the wave-like polarization properties of the turbulent fluctuations, as compared to what linear theory predicts~\citep[see, e.g.,][and Sec.~\ref{sec:spectra}]{BoldyrevAPJ2013}.

In the above context, direct numerical simulations play a key role by providing a controlled test ground for different theories, providing information not accessible to observations.
Enormous efforts have been recently made to understand 3D kinetic turbulence via numerical experiments~\citep[e.g.,][]{HowesPRL2008,GaryAPJ2012,TenBargeAPJL2013,VasquezAPJ2014,ServidioJPP2015,ToldPRL2015,WanPRL2015,WanPOP2016,BanonNavarroPRL2016,ComiselAnGeo2016,CerriAPJL2017,CerriAPJL2018,HughesAPJ2017,KobayashiAPJ2017,HughesAPJL2017,FranciAPJ2018,Franci2018a,GroseljPRL2018,ArzamasskiyAPJ2019,RoytershteynApJ2019,ZhdankinPRL2019,GroseljPRX2019}.
In this Perspective, we combine data from our recent 3D3V studies~\citep{CerriAPJL2017,FranciAPJ2018,GroseljPRX2019} to investigate whether common turbulence features exist in all three 
independently performed simulations (Sec.~\ref{sec:sims}), thus indicating a certain ``universality'' of kinetic-scale turbulence.
Moreover, we also highlight possible model-dependent differences between the 3D hybrid-kinetic and fully kinetic
simulations. 
We mention that this approach follows the general idea of adopting different models (and/or implementations) to study turbulent heating and dissipation in collisionless plasmas that was initiated within the ``Turbulent Dissipation Challange'' framework~\citep{ParasharJPP2015}.
Here we extend similar comparative analysis of the spectral properties that have been previously performed in a reduced two-dimensional setup~\citep[see][]{CerriJPP2017,FranciAPJL2017,GroseljAPJ2017} to the more realistic three-dimensional geometry (Sec.~\ref{sec:spectra}), and we present a new analysis of our data based on local structure functions (Sec.~\ref{sec:S2}). Finally, we discuss possible implications for sub-ion-scale turbulence and future directions emerging from this study (Sec.~\ref{sec:conclusions}).

\section{Data sets}\label{sec:sims}

\begin{figure*}[]
\centering
\includegraphics[width=1.0\columnwidth]{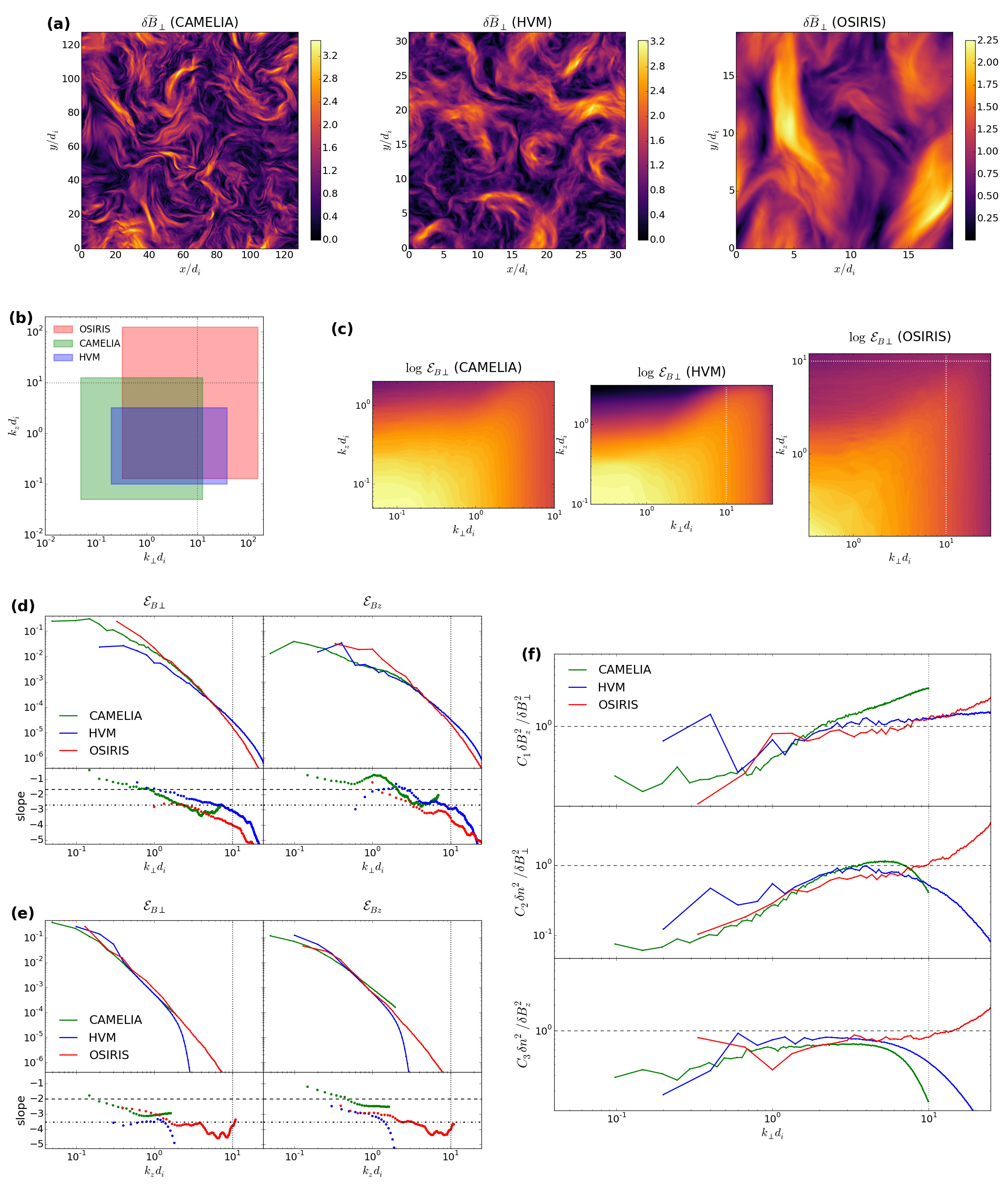}%
\caption{\label{fig1} 
{\bf(a)} $\delta\widetilde{B}_\perp=\delta{B}_\perp/\delta{B}_\perp^{\rm(rms)}$ in a plane perpendicular to $\bb{B}_0$. 
{\bf(b)} Nominal wavenumber-space representation of simulations. 
{\bf(c)} $\delta{B}_\perp$ spectrum in ($k_\perp,\,k_z$) space. 
White dotted lines mark $kd_e=1$. 
{\bf(d)} Top panels: reduced spectra versus $k_\perp d_i$. Spectra have been shifted (see text). Bottom panels: local spectral exponents.
Horizontal lines denote $-5/3$ (dashed) and $-8/3$ (dash-dotted) slopes. Vertical dotted line marks $k_\perp{d_e}=1$. 
{\bf(e)} Same as (e), but versus $k_zd_i$. Here $-2$ (dashed) and $-7/2$ (dash-dotted) slopes are marked for reference.
{\bf(f)} Spectral ratios versus $k_\perp$, normalized to the asymptotic KAW 
prediction (dashed horizontal lines; see text for details).}
\end{figure*}

In the following, we consider three recent kinetic simulations in a six-dimensional phase space (``3D3V'') using: (i) CAMELIA, a hybrid particle-in-cell (PIC) code with massless electrons~\citep{Franci2018a},
(ii) HVM , an Eulerian hybrid-Vlasov code with finite electron-inertia effects~\citep{ValentiniJCP2007},
and (iii) OSIRIS, a fully kinetic PIC code~\citep{Fonseca2002,Fonseca2013}.
Unless otherwise specified, parallel ($\|$) and perpendicular ($\perp$) directions are defined with respect to the global 
mean magnetic field $\bb{B}_0=B_0\bb{e}_z$. 
\citet{FranciAPJ2018} employed the CAMELIA code to investigate freely decaying,
Alfv\'enic fluctuations in a cubic box ($L_\|=L_\perp=128d_i$ with $512^3$ grid points and $2048$ particles per cell (ppc)) for $\beta_i=\beta_e=0.5$, where $\beta_s=8\pi{n_0}T_{s}/B_0^2$ is the species beta.
\citet{CerriAPJL2017} instead adopted the HVM code to study freely decaying compressive fluctuations 
in an elongated box ($L_\|=2L_\perp\simeq63d_i$ with $384^2\times64$ grid points in real space, and $51^3$ points in a velocity space bounded by $|v/v_{th,i}|\leq5$)
for $\beta_i=\beta_e=1$ and with a reduced ion-electron mass ratio of $m_i/m_e=100$ 
(viz. including $d_e$-effects in a generalized Ohm's law).
Spectral filters were applied at runtime, determining a cutoff in the turbulent spectrum at $k_\perp{d_i}>20$ and at $k_zd_i>2$.
Finally, \citet{GroseljPRX2019} use the OSIRIS code to investigate 
continuously driven Alfv\'enic fluctuations in a $\beta_i\approx\beta_e\approx0.5$ plasma with $m_i/m_e=100$. An elongated box was used ($L_\|=2.56L_\perp\simeq48.3d_i$ with $928^2\times1920$ grid points and $150$ ppc per species).
An example of $\delta\widetilde{B}_\perp=\delta{B}_\perp/\delta{B}_\perp^{\rm(rms)}$ in a two-dimensional cut perpendicular to $\bb{B}_0$ is given in Fig.~\ref{fig1}(a), along with a schematic representation of these simulations in the ($k_\perp,\,k_\|$) plane (Fig.~\ref{fig1}(b)).

In the following, the analysis of freely decaying simulations (viz., CAMELIA and HVM) is performed at the peak of the turbulent 
activity~\citep[cf., e.g.,][]{ServidioJPP2015}, while for the continuously driven OSIRIS run we consider the turbulence at the end of the simulation when the kinetic range spectra appear converged.
Following \citet{FranciAPJ2018} and \citet{GroseljPRX2019}, 
PIC data have been filtered before performing the analysis to remove spectral regions dominated by
particle noise.
The OSIRIS data have been filtered for $k_\perp{d_i}>30$ or $k_zd_i>12$ and downsampled to a grid 
$464^2\times640$. 
Note that OSIRIS simulations require to resolve the Debye scale, while the physical scales of interest are well represented at a lower resolution.
A short-time average over $\Delta{t}\Omega_{ce}=2$ ($\Omega_{ce}$ being the electron cyclotron frequency) was also performed to further reduce electron-scale noise~\citep{GroseljPRX2019}.
The CAMELIA data have been filtered for $k_\perp{d_i}>10$ or $k_zd_i>2$.
We also considered alternative filtering approaches 
confirming
that our results are not very sensitive to 
such particular choice.

\section{Spectral slopes and normalized field ratios}
\label{sec:spectra}

Here we review and compare the standard set of spectral properties in our independently performed 3D kinetic simulations, namely the slopes of the turbulence power spectra and the spectral field ratios. 
Early theoretical predictions for sub-ion-range turbulence~\citep[e.g.,][]{ChoLazarianAPJL2004,SchekochihinAPJS2009} proposed a spectral scaling $\sim{k_\perp^{-7/3}}$ for the 
magnetic energy spectrum. 
However, solar wind observations typically 
exhibit
much steeper magnetic spectra, namely $\sim{k_\perp^{-2.8}}$~\citep[e.g.,][]{AlexandrovaPRL2009,AlexandrovaSSRv2013,SahraouiPRL2010,ChenJPP2016,KobayashiAPJ2017,SorrisoValvoSP2018}. 
Similar spectral exponents were also reported in recent 3D kinetic simulations~\citep{ToldPRL2015,CerriAPJL2017,CerriAPJL2018,Franci2018a,FranciAPJ2018,GroseljPRL2018,ArzamasskiyAPJ2019,GroseljPRX2019}. 
Recently, refined predictions were proposed 
to explain steeper spectra. 
Those include
intermittency corrections~\citep{BoldyrevPerezAPJL2012}, dissipative effects~\citep{HowesPOP2011,PassotSulemAPJL2015}, 
and reconnection-mediated turbulence~\citep{LoureiroBoldyrevAPJ2017, MalletJPP2017}. Further insight into the nature of kinetic-scale turbulence can
be obtained from the spectral field ratios, which have been used to detect
wave-like polarization properties in
solar-wind turbulence and
in kinetic simulations~\citep[e.g.,][]{SahraouiPRL2009,SalemAPJL2012,TenBargeApJ2012,ChenPRL2013,KiyaniApJ2013,CerriAPJL2017,FranciAPJ2018,GroseljPRL2018}. 

In Fig.~\ref{fig1}(c) the two-dimensional Fourier spectra, $\mathcal{E}(k_\perp,k_z)$, are shown. The wavenumber region 
($k_\perp,\,k_z$)
occupied by the turbulent fluctuations already highlights the anisotropic nature of the cascade, 
with energy preferentially flowing
to high $k_\perp$. 
However, note that the 2D Fourier spectrum may exhibit a weaker anisotropy than the one typical of turbulent eddies, which are elongated along the \emph{local} field direction~\citep[see, e.g.,][]{ChoVishniacAPJ2000}. 
We perform a local analysis of anisotropy in Sec.~\ref{subsec:anisotropy}.

In Fig.~\ref{fig1}(d)--(e), the reduced 1D spectra, $\mathcal{E}(k_\perp)$ (upper panels), and their local slope (lower panels) are reported. 
To remove the effects of different energy injection conditions, the $k_\perp$-spectra have been normalized so that they overlap in the sub-ion range, at $k_\perp{d_i}\simeq5$. 
According to the spectral anisotropy in Fig.~\ref{fig1}(c), the $k_z$-spectra have been consequently matched at $k_zd_i\simeq0.5$. For our choice of low-pass filter (see Sec.~\ref{sec:sims}),
CAMELIA spectra artificially flatten 
beyond $k_\perp{d_i}\gtrsim7$ due to PIC noise, and therefore we do not show CAMELIA data in
this range in Fig.~\ref{fig1}(d).
Overall, the spectral slopes are consistent with each other, although the spectra 
obtained from the three simulations do not quite assume a universal shape. 
Close to the box scale, the spectral exponents are likely 
affected by the turbulence injection details.  
It is also possible that some of the sub-ion scale spectral exponents are not fully converged in terms of the box size (which was different for each simulation) and of the limited extent of sub-ion range itself. 3D3V simulations with a significantly larger sub-ion range are required to clarify this point.
To some degree, differences at sub-ion scales could also be physical.
In particular, the HVM simulation includes electron inertia
effects in Ohm's law, while the OSIRIS results include the full electron kinetics, 
such as electron Landau damping and finite electron 
Larmor radius corrections. 
It is interesting to notice that OSIRIS 
spectra become steeper than the hybrid counterparts
beyond $k_\perp{d_i}\gtrsim3$, for our particular choice of the mass ratio ($m_i/m_e=100$).
This feature has been usually explained in terms of electron Landau damping~\citep{GroseljAPJ2017,ChenNATCo2019}, which is not included in the hybrid-kinetic model.

In Fig.~\ref{fig1}(f), we report the comparison of spectral ratios, $C_1\,\delta{B_z^2}/\delta{B_\perp^2}$ (top), $C_2\delta{n^2}/\delta B_\perp^2$ (middle), and $C_3\,\delta{n^2}/\delta{B_z^2}$ (bottom). The ratios 
are normalized to the $\beta$-dependent kinetic Alfv\' en wave (KAW) 
eigenvalue from asymptotic linear theory ($\rho_i^{-1}\ll{k_\perp}\ll\rho_e^{-1}$ and $k_\|\ll{k_\perp}$), namely $C_1=(2+\beta)/\beta$, $C_2=(2+\beta)\beta/4$, and $C_3=\beta^2/4$,  
where $\beta=\beta_i+\beta_e$~\citep[see, e.g.,][for details]{BoldyrevAPJ2013}. 
In the normalized units, asymptotic KAW theory predicts a value of unity for all three 
ratios. This is essentially the result of KAWs developing a certain degree of magnetic compressibility at sub-ion scales, which sets the relation between $\delta B_\perp$ and $\delta B_\|$, and requiring that compressive magnetic fluctuations are pressure balanced, which in turn provides a relation between $\delta B_\|$ and $\delta n$~\citep[see, e.g.,][]{SchekochihinAPJS2009,BoldyrevAPJ2013}.
As found in previous studies~\citep[e.g.,][]{SalemAPJL2012,TenBargeApJ2012,ChenPRL2013,CerriAPJL2017,FranciAPJ2018,GroseljPRL2018}, the spectral field ratios are overall consistent with KAW-like turbulence at sub-ion scales.
This is not completely surprising, as both in CAMELIA and OSIRIS simulations, Alfv\'enic fluctuations are injected.
On the other hand, compressible magnetic fluctuations (i.e., including $\delta{B_\|}$) are injected in HVM run, and yet KAW-like fluctuations still develop.
It was also proposed that KAWs may, quite generally, emerge as a result of Alfv\'en waves interacting with large-scale inhomogeneities~\citep{PucciJGR2016}.
Thus, the KAW-like spectral properties at sub-ion scales appear to be a relatively robust feature, independent of the details of the turbulent fluctuations injected at the MHD scales~\citep[cf.][]{CerriJPP2017}.  
While the results are overall consistent, some differences are also seen, most notably in the high-$k_\perp$ 
range ($k_\perp{d_i}\gtrsim10$), which could be presumably attributed to various numerical
artifacts. 
However, some deviations could also relate to 
differences between the
hybrid-kinetic and fully kinetic model (for instance, some 
dispersion relation properties
not being exactly the same~\citep[e.g.,][]{ToldNJP2016}).

So are the sub-ion-scale field polarizations indeed KAW-like?
As discussed above, recent observations and kinetic simulations are consistent with such idea, although linear wave predictions are not necessarily satisfied precisely~\citep[e.g.,][]{KiyaniApJ2013,ChenPRL2013,CerriAPJL2017,FranciAPJ2018}.
\citet{ChenPRL2013} report an average value of 0.75 for the normalized ratio $C_2\delta{n^2}/\delta{B_\perp^2}$, whereas (asymptotic) KAW theory predicts a value of unity. 
That latter may be due to different reasons, among which we remark the following two: (i) sub-ion-range turbulence is not made of purely KAW-like fluctuations, and/or (ii) the asymptotic conditions that are used in the derivation of linear theory predictions are not met exactly because of the limited sub-ion range of scales and/or because of the inherently nonlinear dynamics of turbulence. 
These two explanations are not mutually exclusive, of course.
Indeed, sub-ion-scale turbulence can in principle include contributions from wave-like fluctuations of other nature.
This may include fluctuations consistent with whistler~\citep[e.g.,][]{GaryJGR2009}, ion-cyclotron~\citep[e.g.,][]{OmidiJGR2014,ZhaoJGR2018}, or ion Bernstein waves~\citep[e.g.,][]{PodestaJGR2012,GroseljAPJ2017,DelSarto2017}, to name a few. 
On the other hand, the spectral ratios could also deviate from linear KAW predictions as a result of nonlinear dynamics.
For example, \citet{BoldyrevAPJ2013} propose that, specifically the (normalized) $C_2\delta{n^2}/\delta{B_\perp^2}$ ratio may fall somewhat below the KAW prediction due to a (yet to be investigated) nonlinear effect, analogous to the residual-energy phenomenon in MHD turbulence.

\section{Multi-point structure functions}\label{sec:S2}

\begin{figure*}[]
\centering
\includegraphics[width=0.97\columnwidth]{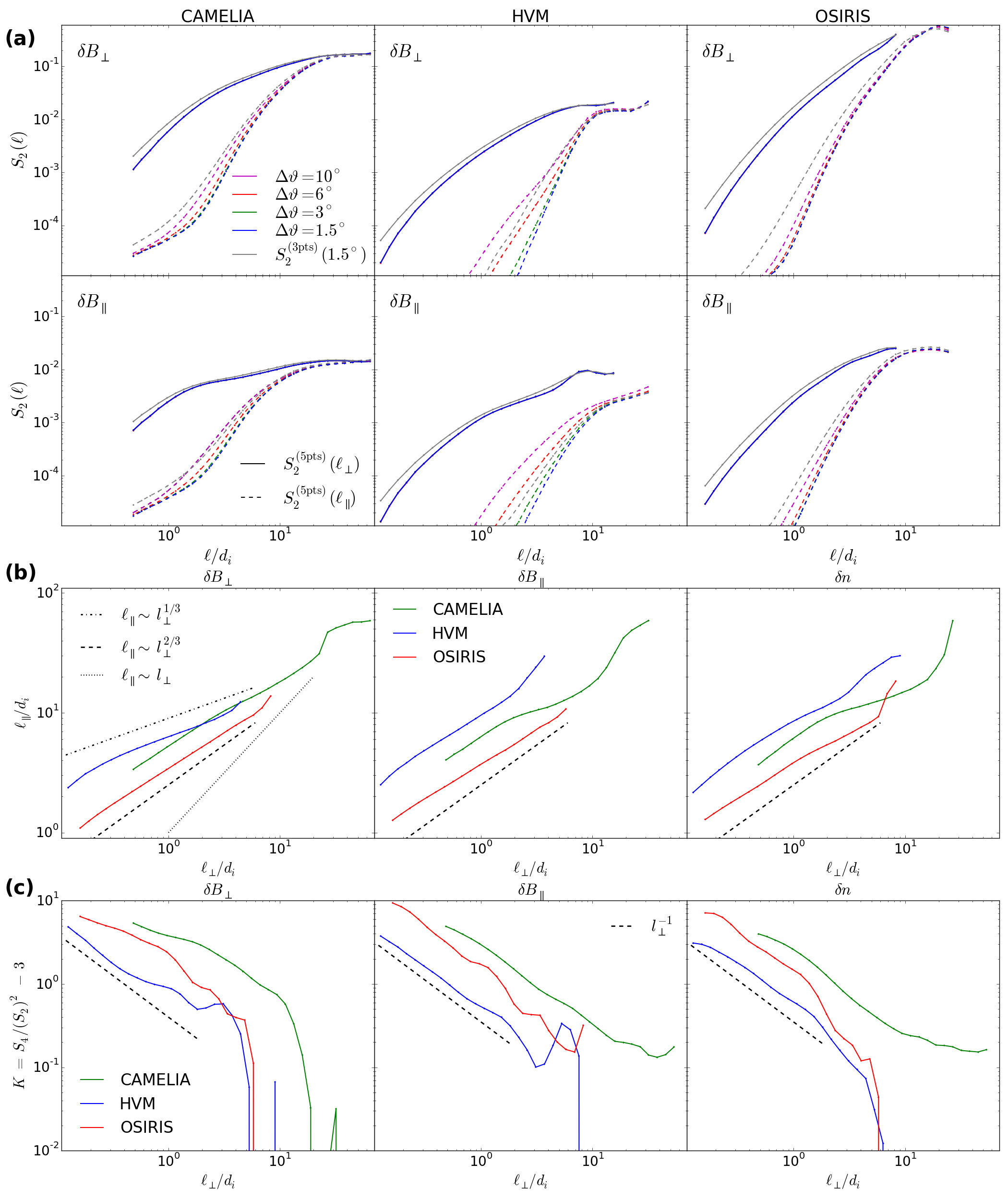}%
\vspace{-0.55cm}
\caption{\label{fig2}
{\bf(a)} Five-points, 2nd-order structure functions, $S_2^{\rm(5)}$, of $\delta{B_\perp}$ (top row) and $\delta{B_\|}$ (bottom row) versus $\ell_\perp$ (continuous lines) and $\ell_\|$ (dashed lines). Here, $\|$ and $\perp$ are defined with respect to the local field direction (i.e., $\delta{B_\|}\neq\delta{B_z}$) with different angular tolerance, $\Delta\vartheta$ (colored lines; see text for definition). 
$S_2^{\rm(3)}$ with $\Delta\vartheta=1.5^\circ$ are also shown for reference (grey lines).
{\bf(b)} Anisotropy scaling, $\ell_\|(\ell_\perp)$, of $\delta{B_\perp}$ (left panel), $\delta{B_\|}$ (central panel), and $\delta{n}$ (right panel), derived from $S_2^{\rm(5)}$ with $\Delta\vartheta_\perp=1.5^\circ$ nominal resolution. Three reference scalings are also shown.
{\bf(c)}  Excess kurtosis,
$K=S_4/[S_2]^2-3$ versus $\ell_\perp$ for $\delta{B_\perp}$ (left panel), $\delta{B_\|}$ (central panel), and $\delta{n}$ (right panel). A $1/\ell_\perp$ scaling is given for reference.}
\end{figure*}

Beyond energy spectra, fluctuations across different scales may be
investigated in more detail via structure 
functions, i.e., the moments of local field increments~\cite[e.g.,][]{FrischBOOK1995,BiskampBOOK2008}.
Two-point structure functions, $S_m^{\rm(2)}$ ($m$ being the order), are most common. However, these 
cannot quantitatively produce
the correct scaling for fluctuations with power spectra steeper than $\sim{k^{-3}}$, assuming a clean power-law spectrum~\citep{Falcon2007,ChoLazarianAPJ2009}. 
Therefore, structure functions using more than two points are generally required at kinetic scales. 
Essentially, higher-order increments yield a scale decomposition that 
is more effective in filtering out the large-scale fluctuations 
below $k\approx \pi/\ell$ in spectral space, where $\ell$ is the increment scale.
We also mention 
that if the signal is a polynomial of degree $\mathcal{N}-2$, its corresponding
2nd-order, $\mathcal{N}$-point structure function vanishes ~\citep{ChoAPJ2019}.
This makes multi-point structure functions more suitable for the analysis of relatively
smooth signals with steep spectra \citep{Schneider2004}.
A detailed review of $\mathcal{N}$-point increments, as well as their physical interpretation 
can be found in \citet{ChoAPJ2019}.
Here, we consider for some field $f(\bb{x})$ the 
\emph{conditional}, five-point structure functions:
\begin{align}
    S_{m}^{(5)}(\ell,\vartheta_{B_{\rm loc}}) = \left\langle|\Delta{f}(\bb{x},\boldsymbol{\ell})|^m\,|
    \ell,\vartheta_{B_{\rm loc}}\right\rangle_{\bb{x}}
\end{align}
where $\Delta{f(\bb{x},{\boldsymbol\ell})}=[f(\bb{x}\!+\!2\boldsymbol{\ell})-4f(\bb{x}\!+\!\boldsymbol{\ell})+6f(\bb{x})-4f(\bb{x}\!-\!\boldsymbol{\ell})+f(\bb{x}\!-\!2\boldsymbol{\ell})]/\sqrt{35}$ is the (normalized) field increment, $\langle\dots\rangle_{\bb{x}}$ is a space average, and $\vartheta_{B_{\mathrm{loc}}}$ is the angle between the increment vector $\boldsymbol{\ell}$ and the local mean magnetic field $\bb{B}_{\rm loc}$. 
The term ``conditional'' implies that $S_m$ are defined as conditional averages of 
$|\Delta{f}(\bb{x},\boldsymbol{\ell})|^m$, using only those points in the statistical sample that fall within a given (narrow) 
range for $\ell$ and $\vartheta_{B_{\rm loc}}$.
We also considered three-point structure functions 
(see Fig.~\ref{fig2}(a)) and, for a limited number of cases,
seven-point structure functions (not shown). 
Comparison between the three-point, five-point and seven-point structure functions shows not only qualitative similarities among the three cases, but an apparent quantitative convergence with increasing number of points. 
We chose to illustrate the results in Fig.~\ref{fig2} in terms of five-point structure functions in order to provide better constraints for the theoretical predictions.
Similar to two-point increments, where the local mean field is often defined as $\bb{B}_{\rm loc}({\bf x},\boldsymbol{\ell})=[\bb{B}(\bb{x})+\bb{B}(\bb{x}\!+\!\boldsymbol{\ell})]/2$~\citep[e.g.,][]{ChoVishniacAPJ2000,Mallet2016}, we obtain ${\bf B}_{\rm loc}$ by averaging over the points used for the increment. For five-point increments, a reasonable definition is
$\bb{B}_{\mathrm{loc}}(\bb{x},\boldsymbol{\ell})=[\bb{B}(\bb{x}\!+\!2\boldsymbol{\ell})+4\bb{B}(\bb{x}\!+\!\boldsymbol{\ell})+6\bb{B}(\bb{x})+4\bb{B}(\bb{x}\!-\!\boldsymbol{\ell})+\bb{B}(\bb{x}\!-\!2\boldsymbol{\ell})]/16$. 
It is straightforward to check that such mean field definition filters out fluctuations around the scale of the increment $\sim\ell$, while preserving the contribution from scales larger than $\ell$.

In what follows, we consider field-perpendicular, $S_m(\ell_\perp)\equiv S_m^{(5)}(\ell_\perp,90^\circ-\Delta\vartheta\!\leq\!\vartheta_{B_{\mathrm{loc}}}\!\leq\!90^\circ)$, and field-parallel, $S_m(\ell_\parallel)\equiv S_m^{(5)}(\ell_\parallel,\vartheta_{B_{\mathrm{loc}}}\!\leq\!\Delta\vartheta)$, five-point structure functions of the magnetic field and density fluctuations, where $\Delta\vartheta$ represents a finite angular tolerance used in practice to determine the \emph{local} perpendicular and parallel directions.
We reduce $\Delta\vartheta$ until the scalings appear converged.
The field increments, from which we obtain the conditional structure functions, are evaluated at every grid point. 
In each grid point and at every scale, increments are sampled along random directions.
The numbers of these random directions per grid point have
been tested to provide a statistically significant (i.e., converged) sample.
The sample that is used in the following is such that any structure 
function $S_m(\ell,\vartheta_{B_{\rm loc}})$ counts {\em at least} $1.5\times10^5$ points 
per scale $\ell$, in any given band for $\vartheta_{B_{\mathrm{loc}}}$.

\subsection{Spectral anisotropy}\label{subsec:anisotropy}

A delicate point concerns the sub-ion-range spectral anisotropy, $k_\|\sim{k_\perp}^{\alpha}$~\citep[cf., e.g.,][]{SchekochihinAPJS2009,BoldyrevPerezAPJL2012,CerriAPJL2018,Landi_arxiv2019}. 
As is known from MHD, electron-MHD (EMHD), and kinetic-reduced-MHD (KRMHD) turbulence~\citep{ChoVishniacAPJ2000,ChoLazarianAPJ2009,MeyrandPNAS2019}, the true anisotropy is often revealed only when measured with respect to the \emph{local}, scale-dependent mean magnetic-field direction. 
Somewhat contradicting estimates, obtained with different methods, for the sub-ion-scale anisotropy have been presented in recent works. 
Here, we revisit this issue using the above-mentioned implementation of five-point structure functions, consistently applied to all data.

In Fig.~\ref{fig2}(a) we show the perpendicular and parallel second-order structure function scalings, and in Fig.~\ref{fig2}(b) we show the inferred anisotropy, $\ell_\|(\ell_\perp)$.
The characteristic parallel length $\ell_\|(\ell_\perp)$ at a given perpendicular scale $\ell_\perp$ is obtained by finding the value of $\ell_\|$, at which the amplitudes of $S_2(\ell_\|)$ and $S_2(\ell_\perp)$ match.
To illustrate the sensitivity to the local mean field direction, we show in Fig.~\ref{fig2}(a) the convergence with respect to the angular tolerance $\Delta\vartheta$.
The parallel scalings appear converged at $\Delta\vartheta\simeq3^\circ$
for CAMELIA data and at around $\Delta\vartheta\simeq1.5^\circ$ for HVM,
whereas the OSIRIS results are somewhat less sensitive to $\Delta\vartheta$ (converging already for $\Delta\vartheta\simeq6^\circ$).
This difference may occur because OSIRIS simulation exhibits the weakest anisotropy (in absolute values). 
Physically, $\Delta\vartheta$ should be approximately no larger than $\sim\ell_\perp/\ell_\|$ of the small-scale turbulent eddies. 
Thus, smaller $\Delta\vartheta$ are needed if a stronger anisotropy develops at the energy-containing scales.

All quantities seem to converge to a scaling close to $\ell_\|\sim{\ell_\perp}^{2/3}$ (although $\delta{B_\perp}$ fluctuations in HVM exhibit a scaling closer to $1/3$ over the range of scales across $\ell_\perp\sim{d_i}(=\rho_i)$).
It is worth noticing, however, that this is not the end of the story, as the scaling is not quite $2/3$ and additional effects such as $B$-field curvature may slightly change the anisotropy. 
Indeed, the field increments are taken along a straight line. 
If the local magnetic field lines are significantly curved over the extent of the increment stencil ($=4\ell$ for five-point increments), the field increments will mix contributions from different field lines, in which case the anisotropy may be somewhat underestimated.
It is worth mentioning that a scaling $\ell_\|\sim{\ell_\perp}^{2/3}$  was proposed in \citet{BoldyrevPerezAPJL2012}, based on a filling-factor correction for the fluctuation energy. 
Assuming the energy is concentrated in intermittent, two-dimensional structures as in \citet{BoldyrevPerezAPJL2012}, the filling factor should scale as $k_\perp^{-1}\sim{l_\perp}$.
The filling factor may be approximately estimated from the inverse scaling of the excess kurtosis~(\citet{MatthaeusRSPTA2015}; see Sec.~\ref{subsec:kurosis}). 
Our results shown in Fig.~\ref{fig2}(c) are indeed roughly consistent with an excess kurtosis scaling $\sim\ell_\perp^{-1}$, although this approximate scaling is overall better satisfied for $\delta{B_\parallel}$ and $\delta{n}$ than for $\delta{B_\perp}$.
Finally, we mention that an alternative anisotropy estimate, based on a spectral band-pass filter~\citep{ChoLazarianAPJ2009}, gives a somewhat stronger anisotropy than the five-point structure functions (not shown). 
On the other hand, qualitatively similar results are still obtained for all data. 
Thus, all simulations analyzed exhibit a similar sub-ion-scale anisotropy according to the particular diagnostics employed. 
Therefore, the differences that were previously reported in the literature could be mainly related to the different methods employed.

\subsection{Intermittency: the ``saturation problem''}\label{subsec:kurosis}

Another relevant feature of kinetic plasma turbulence is the excess kurtosis of the fluctuations, $K(\ell_\perp)=S_4(\ell)/[S_2(\ell)]^2-3$. 
The increase of $K(\ell_\perp)$ above zero is a measure of non-Gaussian statistics of the turbulent fluctuations~\citep{FrischBOOK1995,MatthaeusRSPTA2015}. 
As seen in Fig.~\ref{fig2}(c), the excess kurtosis gradually increases above the Gaussian value throughout the sub-ion scale range.
Moreover, similar statistical trends are seen for $\delta{B_\perp}$, $\delta{B_\parallel}$, and $\delta{n}$
[note that we take here the component of $\delta{\bf B}_\perp$ parallel to $\boldsymbol{\ell}\times {\bf B}_{\rm loc}$
to estimate the flatness of $\delta B_\perp$
\citep[see also][]{KiyaniApJ2013}].
In apparent contrast with our results, a number of observational studies of solar wind turbulence find non-Gaussian, yet nearly \emph{scale-independent} turbulence statistics at sub-ion scales~\citep{Kiyani2009,WuAPJL2013,KiyaniApJ2013,Chen2014}.
Thus, it appears a process operates in the solar wind that saturates the turbulence statistics already near the transition to sub-ion scales ($\ell_\perp\lesssim{d_i}$).
What could be the reason for this apparent contradiction? One clear difference is that the solar-wind fluctuations are already heavily non-Gaussian at MHD scales \citep{Salem2009}, whereas our 3D kinetic simulations do not quite share the same feature due to the limited simulation domain. 
We mention that even large-size 2D kinetic simulations~\citep[e.g.,][]{WanPRL2012,FranciAPJ2015,LeonardisPOP2016} did not yet achieve $K(\ell_\perp)\gg1$ in the MHD range ($\ell_\perp\gg{d_i}$).
In this context, it may be worth pointing out that intermittency in MHD turbulence is commonly associated with the emergence of sheetlike structures~\citep[e.g.,][]{MatthaeusRSPTA2015,Chandran2015,Mallet2017a}, which may break apart via
the tearing instability (causing the field lines to reconnect), once their perpendicular 
aspect ratio exceeds a critical 
threshold~\citep{MatthaeusLamkinPOF1986,MalletMNRAS2017,Boldyrev2017}. 
For sub-ion-scale turbulence, the possible role of magnetic reconnection has been as well highlighted in a number of recent works~\citep[e.g.,][]{FranciAPJ2016,CerriCalifanoNJP2017,FranciAPJL2017,LoureiroBoldyrevAPJ2017,MalletJPP2017,PapiniAPJ2019}.
Moreover, a recent observational study~\citep{VechAPJL2018} argued that the spectral break at the tail of the MHD cascade may be controlled by reconnection.
Therefore, the phenomenology of the cascade may critically depend on the morphology of the intermittent structures at the transition into the kinetic range~\citep{MalletJPP2017}. 
If the structures are indeed sufficiently sheetlike to be tearing unstable, collisionless reconnection might be one possible process that limits the growth of the sub-ion scale kurtosis~\citep[see also][]{Biskamp1990,Chen2014}. 
However, alternative possibilities such as collisionless damping of the fluctuations cannot be ruled out at this stage.

\section{Concluding remarks}\label{sec:conclusions}

So, what is the nature of sub-ion-scale fluctuations? 
From our independently performed 3D3V (hybrid and fully) kinetic simulations, a picture consistent with KAW turbulence phenomenology emerges. 
Moreover, our results imply a scale-dependent anisotropy, together with intermittent statistics of magnetic and density fluctuations at sub-ion scales.
Thus, we conclude that within the range of parameters explored here, the statistical properties of ion-scale plasma turbulence (at $\beta\sim1$) definitely show a certain degree of similarity, regardless of the precise details of the large-scale energy injection.
On the other hand, slight differences can also be identified, some of which may be also model-dependent.

A number of key aspects will have to await the next-generation of 3D3V kinetic simulations. 
Ideally, future numerical experiments should aim to resolve both larger (MHD) scales, as well as a broader range between the ion and the electron scales by adopting significantly higher (if not realistic) mass ratios. 
These two aspects indeed appear to be both required in order to achieve (i) a possible saturation of the kurtosis at ion scales and (ii) a relevant sub-ion range of scales before electron-scale effects significantly come into play.
Moreover, different aspects other than the spectral and statistical properties of the turbulent fluctuations will need to be considered in characterizing kinetic-range turbulence, as for instance, the dissipation mechanisms of turbulent fluctuations under different plasma conditions and the consequent energy partition among different species~\citep[e.g.,][]{MatthaeusAPJL2016,ParasharAPJL2018,ArzamasskiyAPJ2019,KawazuraPNAS2019,ZhdankinPRL2019}.

While certain progress was definitely achieved in recent years, many other plasma regimes and setups may need to be explored, and the process(es) underlying a possible universality of kinetic-range plasma turbulence (e.g., magnetic reconnection) need to be fully worked out.
Moreover, a few relevant discrepancies between numerical simulations, theories and \emph{in-situ} observations appear.
These ``anomalies'' definitely call for an explanation by the space physics community.
In this context, advances cannot be achieved without investing in next-generation \emph{multi-spacecraft} missions. 
Multi-point \emph{in situ} measurements of turbulent fluctuations from a large number of spacecrafts are indeed fundamental in order to disentangle the nonlinear spatio-temporal character of plasma turbulence~\citep[see, e.g.,][]{Klein_Plasma2020Decadal,Matthaeus_Plasma2020Decadal,TenBarge_Plasma2020Decadal}. 
This includes answering fundamental questions about, for instance, (i) the distribution of turbulent energy in space and time, (ii) the three-dimensional anisotropic structure of energy transfer across scales, (iii) the high-order statistics of the fluctuations, and (iv) the validity of Taylor's hypothesis over a broad range of time and spatial scales.
Alongside observations, advances in computational capabilities are required to perform more realistic numerical simulations as discussed above, and compare these with spacecraft measurements.
Finally, following the same spirit promoted by the ``Turbulent Dissipation Challange''~\citep{ParasharJPP2015}, we would like to end this Perspective by stressing that our community could benefit from comparisons such as the one performed here, involving various codes, models and diagnostics.

\emph{Note added:} \citet{ArzamasskiyAPJ2019} recently reported a scale-independent 
anisotropy at ion scales (i.e., $\ell_\parallel\sim \ell_\perp$) 
based on a set of 3D driven hybrid-kinetic turbulence simulations. 
Using our structure function diagnostic applied to their data, we were able to qualitatively (and quantitatively) reproduce their result. 
A more detailed investigation along these lines is currently ongoing, but beyond the scope of this Perspective and will be presented elsewhere.

\section*{Conflict of Interest Statement}
The authors declare that the research was conducted in the absence of any commercial or financial relationships that could be construed as a potential conflict of interest.

\section*{Author Contributions}
SSC, DG, and LF provided their HVM, OSIRIS and CAMELIA simulation data, respectively. SSC performed the spectral analysis, produced the figures, and wrote the paper, taking into account suggestions from DG and LF. DG implemented and performed the structure function analysis. All authors discussed the results.

\section*{Funding}
SSC is supported by the National Aeronautics and Space Administration under Grant No.~NNX16AK09G issued through the Heliophysics Supporting Research Program.
LF is supported by the UK Science and Technology Facilities Council (STFC) grant ST/P000622/1.

\section*{Acknowledgments}

SSC and DG acknowledge the generous hospitality of the Wolfgang Pauli Institute in Vienna, where the first discussions leading to this work took place.
We acknowledge PRACE for awarding us access to Marconi at CINECA, Italy, where the calculations with the HVM code were performed under the grant No.~2017174107. 
The Cray XC40, Shaheen, at the King Abdullah University of Science \& Technology (KAUST) in Thuwal, Saudi Arabia was utilized for the simulation performed with the OSIRIS code.
LF acknowledges PRACE for awarding him access to  Cartesius at SURFsara, the Netherlands, through the DECI-13 (Distributed European Computing Initiative) call (project HybTurb3D) where the HPIC simulation was performed, and INAF and CINECA for awarding him access to Marconi within the framework of the MoU ``New Frontiers in Astrophysics: HPC and New Generation Data Exploration'' (project INA17\textunderscore C4A26), where new further analysis of the HPIC data has been performed.
The authors would like to acknowledge the OSIRIS Consortium, consisting of UCLA and IST (Lisbon, Portugal) for the use of OSIRIS 3.0 and for providing access to the OSIRIS 3.0 framework.
SSC acknowledges Dr.~C.~Cavazzoni and Dr.~M.~Guarrasi (CINECA, Italy) for their contributions to HVM code parallelization, performance and implementation on Marconi-KNL.
The authors also acknowledge useful discussions with Alfred Mallet, Lev Arzamasskiy, Bill Dorland, Matt Kunz, Simone Landi, Emanuele Papini, 
Frank Jenko, and David Burgess.

\section*{Data Availability Statement}
The data used in this study are available from the authors upon reasonable request.


\end{document}